\documentclass[10pt]{article}

\usepackage{graphicx} % Required for inserting images

\usepackage[T1]{fontenc}
\usepackage[latin9]{inputenc}

\usepackage[scaled]{helvet}

\usepackage{geometry}
\usepackage{color}
\usepackage{verbatim}
\usepackage{float}
\usepackage{mathtools}
\usepackage{algorithm2e}
\usepackage{amsthm,amsmath,amsfonts}
\usepackage{amssymb}
\usepackage{setspace}
\usepackage{array} 
\usepackage{wasysym}
\usepackage{esint}
\usepackage[round,authoryear]{natbib}
\usepackage[unicode=true,
 bookmarks=false,
 breaklinks=true,pdfborder={0 0 0},pdfborderstyle={},backref=false,colorlinks=true]
 {hyperref}
\hypersetup{citecolor=blue, linkcolor=blue}

\usepackage{bm,dsfont}
\usepackage{cases}
\usepackage[inline]{enumitem}
\usepackage{mathabx} %This package includes "\widecheck"
\makeatletter
\theoremstyle{plain}

\theoremstyle{plain}

\theoremstyle{plain}

\numberwithin{prop}{section}
\theoremstyle{plain}

\numberwithin{thm}{section}
\theoremstyle{plain}

\numberwithin{cor}{section}
\theoremstyle{plain}

\theoremstyle{plain}

\advance \topmargin by -\headheight
\advance \topmargin by -\headsep

\usepackage{booktabs,tabularx,array}
\newcolumntype{Y}{>{\raggedright\arraybackslash}X}
\usepackage{mathrsfs}
\usepackage{graphicx,xcolor}
\usepackage{textcomp}
\usepackage{times}
\usepackage[final]{microtype}
\usepackage{titling}
\LinesNumbered
\RestyleAlgo{algoruled}

\title{Fast Computational Methods for Regularized Estimating Equations}
\date{}

\author{
Weihua Shi\thanks{Equal contribution, Department of Mathematics and Statistics, McGill
University},
Yixuan Li\thanks{Equal contribution, Department of Mathematics and Statistics, McGill
University},
Yi Lian\thanks{Perelman School of Medicine, University of Pennsylvania},
Archer Y. Yang \thanks{Co-corresponding author, Department of Mathematics and Statistics, McGill University; Mila (archer.yang.yi@gmail.com)},
Yue Zhao \thanks{Co-corresponding author, Department of Mathematics, University of York (yue.zhao@york.ac.uk)}
}

\begin{document}
\global\long\def\argmin{\operatorname*{arg\,min}}%

\global\long\def\argmax{\operatorname*{arg\,max}}%

\global\long\def\sgn{\operatorname*{sgn}}%

\global\long\def\bU{\mathbf{U}}%

\global\long\def\cone{\mathscr{C}}%

\global\long\def\cz{\mathscr{Z}}%

\global\long\def\expt{\mathscr{E}}%

\global\long\def\pr{\mathrm{P}}%

\global\long\def\var{\mathrm{var}}%

\global\long\def\E{\mathbb{E}}%

\global\long\def\real{\mathbb{R}}%

\global\long\def\prob{\mathbb{P}}%

\global\long\def\bbX{\mathbb{X}}%

\global\long\def\NN{\mathbb{N}}%

\global\long\def\R{\mathbb{R}}%

\global\long\def\RR{\mathbb{R}}%

\global\long\def\S{S_{\mathcal{G}}}%

\global\long\def\s{s_{\mathcal{G}}}%

\global\long\def\Sp{S_{\mathcal{G}}^{c}}%

\global\long\def\PP{\mathbb{P}}%

\global\long\def\Rn{R_{n}}%

\global\long\def\MnR{\mathbb{M}(\Rn)}%

\global\long\def\MnpR{\mathbb{M}^{+}(\Rn)}%

\global\long\def\T{\top}%

\global\long\def\cmpt{{\scriptscriptstyle \text{c}}}%

\global\long\def\veps{\varepsilon}%

\global\long\def\event{\mathcal{E}}%

\global\long\def\actset{\mathcal{A}}%

\global\long\def\smallo{{\scriptstyle \mathcal{O}}}%

\global\long\def\bigo{\mathcal{O}}%

\global\long\def\calP{\mathcal{P}}%

\global\long\def\calS{\mathcal{S}}%

\global\long\def\calG{\mathcal{G}}%

\global\long\def\calN{\mathcal{N}}%

\global\long\def\calNg{\calN_{g,1/4}}%

\global\long\def\calNgp{\calN_{g',1/4}}%

\global\long\def\orc{{\scriptstyle \text{o}}}%

\global\long\def\initial{{\scriptstyle \text{initial}}}%

\global\long\def\lasso{{\scriptstyle \text{lasso}}}%

\global\long\def\zero{\mathbf{0}}%

\global\long\def\one{\mathbf{1}}%

\global\long\def\ddd{,\ldots,}%

\global\long\def\ind{\mathbb{I}}%

\global\long\def\diag{\operatorname{diag}}%

\global\long\def\prox{\operatorname{prox}}%

\global\long\def\Proj{\operatorname{Proj}}%

\global\long\def\dom{\mathrm{dom}}%

\global\long\def\ku{\kappa_{\mathbf{U}}}%

\global\long\def\tu{\tau_{\mathbf{U}}^{2}}%

\global\long\def\bUn{\bU_{n}}%

\global\long\def\Xij{X_{ij}}%

\global\long\def\eij{\varepsilon_{ij}}%

\global\long\def\eji{\varepsilon_{ji}}%

\global\long\def\Yij{Y_{ij}}%

\global\long\def\Yji{Y_{ji}}%

\global\long\def\piij{{\pi(i),\pi(\lfloor n/2\rfloor+i)}}%

\global\long\def\epiij{\varepsilon_{\piij}}%

\global\long\def\barm{\overline{m}}%

\global\long\def\bfz{\mathbf{0}}%

\global\long\def\Mndelta{\mathbb{M}_{n}(\delta)}%

\global\long\def\bXpiij{ X_{\piij}}%

\global\long\def\sumij{\underset{i\neq j}{\sum\sum}}%

\global\long\def\xii{\xi_{i}}%

\global\long\def\bvg{\bv_{g}}%

\global\long\def\Dn{D_{n}}%

\global\long\def\elln{l_{n}}%

\global\long\def\bhz{\hat{\beta}^{0}}%

\global\long\def\bhi{\hat{\beta}^{\text{initial}}}%

\global\long\def\bs{\beta^{*}}%

\global\long\def\bho{\hat{\beta}^{\text{oracle}}}%

\global\long\def\barg{\overline{g}}%

\global\long\def\card{\operatorname*{card}}%

%----------------------------------------

\newcommand\colorred[1]{{\textcolor{red}{#1}}}
\newcommand\colorm[1]{{\textcolor{magenta}{#1}}}
\newcommand\CG[1]{{\textcolor{green}{#1}}}
\newcommand\colorgray[1]{{\textcolor{gray}{#1}}}
\newcommand\CC[1]{{\textcolor{cyan}{#1}}}

\global\long\def\bx{\mathbf{x}}
\global\long\def\wtvare{\widetilde\varepsilon}
\global\long\def\rmd{{\rm d}}
\def\wh{\widehat}
\def\wt{\widetilde}

%----------------------------------------
%Added by Yue

\global\long\def\bbW{\mathbb{W}}
\global\long\def\bbZ{\mathbb{Z}}
\global\long\def\bbY{\mathbb{Y}}
\global\long\def\Deltac{\Delta_\textup{c}}
\global\long\def\deltac{\delta_\textup{c}}
\global\long\def\calA{\mathcal{A}}
\global\long\def\calF{\mathcal{F}}
\global\long\def\barUn{\bar\mathbf{U}_{n}}
\global\long\def\wtUn{\widetilde{\mathbf{U}}_{n}}
\global\long\def\wtA{\widetilde{A}}
\global\long\def\barSigma{\overline{\Sigma}}
\global\long\def\bX{\mathbf{X}}
\global\long\def\bY{\mathbf{Y}}
\global\long\def\bmu{\bm{\mu}}
\global\long\def\thetaij{\theta_{ij}}
\global\long\def\bXik{\bX_{ik}}
\global\long\def\bXil{\bX_{i\ell}}
\global\long\def\psiik{\psi(\bXik^\top\beta)}
\global\long\def\psiil{\psi(\bXil^\top\beta)}
\global\long\def\dotphi{\dot{\phi}}
\global\long\def\dotphiik{\dotphi(\bXik^\top\beta)}
\global\long\def\dotphiil{\dotphi(\bXil^\top\beta)}
\global\long\def\wtSn{\widetilde{\mathbf{S}}_{n}}
\global\long\def\wtSnok{\widetilde{\mathbf{S}}_{n,1,k}}
\global\long\def\wtSntkl{\widetilde{\mathbf{S}}_{n,2,k,\ell}}
\global\long\def\wtSnhkl{\widetilde{\mathbf{S}}_{n,3,k,\ell}}
\global\long\def\wtbeta{\widetilde{\beta}}
\global\long\def\bXk{\bX_{k}}
\global\long\def\bxk{\bx_{k}}
\global\long\def\bXl{\bX_{\ell}}
\global\long\def\bxl{\bx_{\ell}}
\global\long\def\BB{\mathbb{B}}
\global\long\def\bG{\mathbf{G}}
\global\long\def\bv{\mathbf{v}}
\global\long\def\calA{\mathcal{A}}
\global\long\def\calB{\mathcal{B}}
\global\long\def\calD{\mathcal{D}}
\global\long\def\calE{\mathcal{E}}
\global\long\def\calL{\mathcal{L}}
\global\long\def\calO{\mathcal{O}}
\global\long\def\calT{\mathcal{T}}
\global\long\def\rn{r_n}
\global\long\def\BB{\mathbb{B}}%
\global\long\def\GG{\mathbb{G}}%
\global\long\def\KK{\mathbb{K}}%
\global\long\def\XX{\mathbb{X}}%
\global\long\def\YY{\mathbb{Y}}%
\global\long\def\us{\textup{s}}%
\global\long\def\thetaij{\theta_{ij}}%
\global\long\def\bUn{\mathbf{U}_{n}}
\global\long\def\bA{\mathbf{A}}
\global\long\def\bM{\mathbf{M}}
\global\long\def\barUn{\overline{\mathbf{U}}_{n}}
\global\long\def\Xij{X_{ij}}
\global\long\def\Xijp{X_{ij'}}
\global\long\def\Yij{Y_{ij}}
\global\long\def\Yijp{Y_{ij'}}
\global\long\def\sij{s_{ij}}
\global\long\def\sijp{s_{ij'}}
\global\long\def\whsij{\wh s_{ij}}
\global\long\def\whsijp{\wh s_{ij'}}
\global\long\def\muij{\mu_{ij}}
\global\long\def\muijp{\mu_{ij'}}
\global\long\def\whmuij{\wh\mu_{ij}}
\global\long\def\whmuijp{\wh\mu_{ij'}}
\global\long\def\deltabeta{\delta_\beta}
\global\long\def\betastar{\beta^*}
\newcommand{\squad}{\hspace{0.5em}}

\global\long\def\nc{\textup{nc}}

\global\long\def\calGstar{\calG^*}%
\global\long\def\calGstarc{\calG^{*c}}%

\global\long\def\calG{\mathcal{G}}%
\global\long\def\calGc{\mathcal{G}^{c}}%
\global\long\def\calGbar{\bar\mathcal{G}}%
\global\long\def\op{\textup{op}}%

\global\long\def\calR{\mathcal{R}}%

\global\long\def\calB{\mathcal{B}}

\global\long\def\PPn{\PP_n}%
\global\long\def\PPnjjp{\PP_{n,jj'}}%
\global\long\def\PPjjp{\PP_{jj'}}%
\global\long\def\dn{d_n}
\global\long\def\an{a_n}
\global\long\def\bn{b_n}
\global\long\def\calTbar{\overline{\calT}}

%\AtAppendix{\counterwithin{lem}{section}}
%\AtAppendix{\counterwithin{lemma}{section}}

\maketitle

\begin{abstract}
Estimating equations arise in a wide range of statistical applications, including longitudinal and clustered data analysis, survival analysis, econometrics, and semiparametric inference.  In high-dimensional settings, adding sparsity-inducing regularization often leads to computational challenges that are not fully addressed by standard penalized optimization routines. These challenges are closely tied to the structural form of the underlying estimating problem: mainly, the estimating function needs not be the gradient of a scalar objective and may involve asymmetric Jacobians, overidentification, nonsmoothness, nonconvexity, or nested optimization.

This article first reviews the application areas of estimating equations, and then the computational methods for regularized estimating equations by organizing them into four broad formulations: minimization-type, Dantzig-type, regularization-type, and fixed-point-type approaches. We discuss the main numerical strategies associated with each formulation, including penalized optimization, constrained linear programming, iterative root-solving, and proximal fixed-point iteration. We also highlight the connection between regularized estimating equations and fixed-point problems, which provides a unified computational perspective for analyzing and solving regularized estimating equations.
\end{abstract}


\section{Introduction\label{sec:methodology}}
\label{sec:background}

%In statistics, optimization methods that estimate some target parameters $\beta\in\RR^p$ by the minimizer of some objective function are ubiquitous.  Minimizing an objective function often leads to an equivalent formulation that seeks the root of some appropriate \textit{estimating function}.

In statistics, a common approach to estimating a target parameter $\beta^* \in \RR^p$ is to minimize a scalar objective function.  Under regularity conditions, stationary points of the objective are characterized by the corresponding gradient equation.   For instance, in maximum likelihood, given a smooth log-likelihood $\ell(\beta)$, a stationary point of the objective $-\ell(\beta)$ satisfies the score equation $\nabla_\beta \ell(\beta) = \bfz$.  More broadly, an \textit{estimating function} is a vector-valued function $\bU(\cdot\,;\beta)$ that maps $\beta$ to $\RR^q$ and whose population expectation has a unique and well-isolated root at the target parameter $\beta^*$, so $\E[\bU(W;\beta)] = \bfz$ only at $\beta=\beta^*$; here $W$ denotes a full hypothetical observation for a generic subject, and $W$ could include the response $Y$, the covariates $X$ (typically with dimension $p$), and any auxiliary variables such as treatment indicators or censoring information.  The corresponding root-finding problem is called an \textit{estimating equation}.  In practice, one works with the sample-averaged estimating function typically formed as $\bUn(\beta) = n^{-1}\sum_{i=1}^{n}\bU(W_i;\beta)$ where the observed samples $W_i$, $i=1,\dots,n$ are i.i.d.\,copies of $W$.  (For simplicity we regard $\mathbf U_n(\cdot)$ as a function of $\beta$ only, and omit displaying the dependence on the observations, the $W_i$'s.)  We then estimate $\beta^*$ by the root $\hat\beta$ such that $\bUn(\beta) = \bfz$ at $\beta=\hat\beta$.  Sometimes an exact root does not exist (for instance when $\bUn(\beta)$ is not smooth as in quantile regression), in which case we seek $\hat\beta$ such that $\bUn(\hat\beta) \approx \bfz$.  Our examples later are often stated at the sample level.

Not every estimating function arises as the gradient of a scalar objective.  For instance, the Jacobian of $\bUn$ may be asymmetric, or the system may be overidentified (that is, $q>p$).  In some of these cases an optimization formulation still exists.  However, standard penalized optimization algorithms such as coordinate descent, iterative thresholding, and local linear approximation usually assume that $\bUn$ is the gradient of a scalar objective, and hence strictly speaking their applications are not always justifiable when this structure is absent.  The computational consequences become especially pronounced once a sparsity penalty is added.

The rest of this manuscript is organized as follows. In Section~\ref{sec:EE_examples} we illustrate a range of (regularized) estimating equations that are drawn from several application domains (which are also summarized in Table~\ref{tab:ee_domains}).  In Section~\ref{sec:algorithms} we address the associated computational approaches, and in particular highlight our recent approach based on fixed-point problems.  Section~\ref{sec:conclusion} concludes.

%structural features that give rise to this situation.

%, or $\bUn$ may fail to be continuous
% DOMAIN 1: Longitudinal and clustered data
\section{Examples of estimating equations}
\label{sec:EE_examples}

\subsection{Longitudinal and clustered data}

Estimating equations are widely encountered in the analysis of longitudinal and clustered data, where repeated measurements on the same subject induce within-cluster dependence and specifying a full joint likelihood for the response vector is often impractical or unnecessary.  Building on the quasi-likelihood idea of \citet{wedderburn1974quasi}, \citet{LiangZeger1986} proposed the generalized estimating equation (GEE) framework, which targets the marginal mean structure directly without requiring full distributional specification. For simplicity, we assume that our $n$ subjects share a common cluster size $k$.  Then, the estimating equation in the GEE formulation is
\begin{equation}\label{eq:gee}
\bUn(\beta) \;=\; \frac{1}{n}\sum_{i=1}^{n} \bbX_i^\T A_i^{1/2}(\beta)\, \wh\Sigma^{-1}\, A_i^{-1/2}(\beta) \bigl\{ \mu_i(\beta) - \YY_i \bigr\} \;=\; \bfz,
\end{equation}
where $\YY_i\in\RR^{k}$ collects the $k$ repeated responses of subject $i$, $\bbX_i\in\RR^{k\times p}$ is the associated covariate matrix, $\mu_i(\beta)$ is the marginal mean vector, $A_i(\beta) = \diag\{\operatorname{Var}(Y_{ij}\mid X_{ij})\}_{j=1}^{k}\in\RR^{k\times k}$ is a diagonal matrix of marginal variances, and $\wh\Sigma\in\RR^{k\times k}$ is a working correlation matrix estimated from the data.  Because $\bUn$ in \eqref{eq:gee} is constructed from a working first-moment specification rather than derived as the gradient of a log-likelihood, unless $\wh\Sigma$ is the identity matrix, the Jacobian associated with $\bUn$ is generally asymmetric \citep{Wang2011}.  This precludes representing $\bUn$ as the gradient of a scalar objective.  Extensions to the GEE include second-order generalized estimating equations (GEE2) \citep{prentice1991estimating, liang1992multivariate}, alternating logistic regression (ALR) \citep{carey1993modelling}, quasi-least squares (QLS) \citep{chaganty1999eliminating}, stochastic second-order GEE \citep{chen2020stochastic}, GEE boosting \citep{wang2024generalized}, and quantile regression \citep{ZuLianGreenYu2023}; asymptotic results and model selection criteria were also developed by \citet{Pan2002}, \citet{XieYang2003}, and \citet{BalanSchiopu-Kratina2005}.  When the working correlation is expanded in a basis of $M$ known matrices $\{R_1,\ldots,R_M\}$, the quadratic inference function (QIF) method of \citet{qu2000improving} stacks the $M$ component score contributions into an extended score vector of dimension $Mp$, producing an overidentified system (as $Mp > p$) that is solved by minimizing a quadratic form similar to that used in generalized method of moments (GMM; see Section~\ref{sec:econometrics}, and also \cite{song2009quadratic, offorha2023analysing}).  Adding a sparsity penalty to \eqref{eq:gee} yields a penalized GEE
formulation for variable selection under correlated responses
\citep{16wang2012penalized}. Such developments are especially important in high-dimensional longitudinal settings, where the dimension of the covariates may far exceed the sample size while within-subject dependence must still be accommodated. For example, penalized GEE methods have been developed for genomic and gene-expression studies with $p \gg n$ \citep{22tian2021gee}. Related GEE-based inferential developments have also been studied for high-dimensional proteomic data \citep{xia2022inference}. Beyond high-dimensional covariates, the same estimating equation framework has been extended to neuroimaging settings where covariates may be tensor-valued \citep{zhang2019tensor} or responses exhibit structured dependence across measurement channels, as in event-related potential data \citep{20lin2020group}.
\subsection{Survival and censored data}
% DOMAIN 2: Survival and censored data

While the difficulty in GEE is an asymmetric Jacobian, survival analysis with right censoring 
introduces an additional, qualitatively different obstacle: the estimating function can be 
nonsmooth. Consider the accelerated failure time (AFT) model for subject $i$: 
\(\log T_i = X_i^\T \beta^* + \epsilon_i\), where \(T_i\) is the failure time. The observed data are 
\((\widetilde{T}_i,\delta_i)\equiv(\min(T_i,C_i),\ind(T_i\le C_i))\), \(i=1,\dots,n\), where the \(C_i\)'s are the censoring time.  Rank-based inference can be conducted through the weighted log-rank estimating equation
\begin{equation}\label{eq:aft_rank}
\bUn(\beta)
=
\frac{1}{n}\sum_{i=1}^{n}
\delta_i\,\varphi\{\beta;e_i(\beta)\}
\bigl[
X_i-\bar X\{\beta;e_i(\beta)\}
\bigr]
=
\bfz,
\end{equation}
where \(e_i(\beta)=\log \widetilde T_i-X_i^\T\beta\) is the $i$-th residual, 
\(\varphi\{\beta;t\}\) is a weight function corresponding to choices such as 
Gehan and log-rank weights, and \(\bar X\{\beta;t\}\) is the corresponding 
risk-set covariate average, with both \(\varphi\{\beta;t\}\) and 
\(\bar X\{\beta;t\}\) evaluated at \(t=e_i(\beta)\) in the \(i\)-th summand in \eqref{eq:aft_rank}
\citep{Tsiatis1990,Ritov1990,jin2003rank}.
Because the risk sets depend on the ordering of the residuals \(e_i(\beta)\), which changes 
only when residuals cross, the estimating function is piecewise constant in \(\beta\) and hence 
is discontinuous at the jump boundaries. The relationship between \eqref{eq:aft_rank} 
and optimization also depends on the choice of weight function: with Gehan weights, 
\(\bUn\) is the subgradient of a convex, absolute-value-type loss that can be minimized by 
linear programming (LP). However, this ``inversion'' to a scalar objective does not extend to 
general weights such as the log-rank weight \citep{jin2003rank}. Induced smoothing 
\citep{brown2007induced,chiou2015rank} replaces \(\bUn\) with a smooth approximation 
obtained by averaging over normal perturbations of the coefficients, restoring differentiability 
without requiring bandwidth selection, while the Buckley--James estimator 
\citep{BuckleyJames1979} recasts the problem as iterative imputation of censored responses 
using the Kaplan--Meier estimate of the residual distribution. Penalized extensions 
\citep{47cai2009regularized,45wang2008doubly} must handle the interaction between this 
piecewise-constant structure and the penalty.
\subsection{Efficient estimation in semiparametric models}
% DOMAIN 3: Causal inference
Asymmetric Jacobians also arise in semiparametric models, where
%treatment effect estimation in the presence of high-dimensional confounders requires semiparametric methods
the parameter of interest $\beta$ is accompanied by a nuisance parameter $\eta$.  Such models are common in the areas of treatment effects, missing data and censoring, and transformation models (such as copula models, where the marginal distribution functions often constitute the nuisance parameter); see \cite{bickel1993efficient, robins1994estimation, Tsiatis1990} for concrete examples.  
%such as the propensity score and outcome regression \citep{robins1994estimation, tsiatis2006semiparametric}.
Our ability to estimate $\beta$ is negatively influenced by the presence of the nuisance $\eta$, but \textit{semi}parametrically efficient estimation of $\beta$ can be conducted based on the efficient score function.  Let $\dot\ell_{\beta,\eta}$ denote the ordinary score for $\beta$, and let $\Pi_{\beta,\eta}$ denote the orthogonal projection onto the closure of the linear span of the tangent set for the nuisance $\eta$ in $L_2(P_{\beta,\eta})$ where $P_{\beta,\eta}$ denotes the model at $\beta,\eta$.  Then the efficient score function is given by
\begin{equation}\label{eq:eff_score}
\tilde\ell_{\beta,\eta} \;=\; \dot\ell_{\beta,\eta} \;-\; \Pi_{\beta,\eta}\,\dot\ell_{\beta,\eta} 
\end{equation}
\citep{bickel1993efficient, van2000asymptotic}.  In many interesting applications the efficient scores can be calculated analytically but, because they are obtained through a projection, they may very well not correspond to the gradient in $\beta$ of any scalar objective (see the discussion at the bottom of p.~351 in \cite{kosorok2008introduction}).  Thus, the estimation of $\beta$ is naturally formulated as solving the estimating equation defined by the efficient scores, rather than as a direct optimization problem.  Specifically, the sample-level estimating equation becomes
\begin{equation}\label{eq:eff_score_ee}
\bUn(\beta) \;=\; \frac{1}{n}\sum_{i=1}^{n} \tilde\ell_{\beta,\wh\eta}(W_i) \;=\; \bfz,
\end{equation}
where $\wh\eta$ is an estimate of the nuisance parameter sometimes obtained by machine learning methods \citep{chernozhukov2018double}.  Regularized extensions arise when $p$ is large, and the algorithm must handle both the sparsity penalty on $\beta$ and the dependence on $\wh\eta$.

%depends on both $\beta$ and $\eta$ through the projection in \eqref{eq:eff_score}, so
%, become then a $p$-dimensional root-finding problem in $\beta$ whose solution depends on $\wh\eta$.
%Semiparametric estimating equations with similar structure appear in debiased high-dimensional inference \citep{GautieRose2021, 61NeykovNingLiuLiu2018} and sufficient dimension reduction \citep{DingSuZhuWang2021}.

\subsection{Econometrics}
\label{sec:econometrics}

Overidentification, namely having more (moment) conditions than unknown parameters, is a characteristic feature of many econometric estimation problems and often calls for the generalized method of moments (GMM) method.  A familiar example arises in linear instrumental variable (IV) regression, where the dimension of the instrument $Z\in\RR^q$ may exceed the dimension of the covariates, yielding an overidentified system of the form
$\E[Z(Y-X^\T\beta)] = \mathbf 0$ (at $\beta=\beta^*$).
More generally, given an estimating function $\bUn(\beta)$,
%let $g(\cdot;\beta)\in\mathbb R^q$, with $q\ge p$, denote a criterion function satisfying
%$
%\E[g(W;\beta^*)] = \mathbf 0.
%$
%Writing
%$
%\bar g_n(\beta)=\frac1n\sum_{i=1}^n g(W_i;\beta),
%$
the GMM estimator is obtained by minimizing the quadratic criterion
$
Q_n(\beta) \equiv \bUn(\beta)^\T S_n \bUn(\beta)
$
over $\beta$, for a positive-definite weighting matrix $S_n=S_n(\beta)$ that could depend on $\beta$ \citep{hansen1982large}.  Subsequent work studied alternative implementations of this criterion, including two-step, iterative, and continuously updating estimators --- depending on whether the $\beta$ in $S_n(\beta)$ is fixed at an initial estimate, is updated in an iterative manner with the $\beta$ in $\bUn(\beta)$, or concurrently with the $\beta$ in $\bUn(\beta)$ --- and examined how the construction of the weighting matrix affects finite-sample behavior \citep{hansen1996finite}.
%\colorm{depending on whether and how we update the weighting matrix $S_n$ with $\beta$}

In modern high-dimensional settings, the computational issue is no longer only overidentification itself, but also how to regularize or select among many candidate moment conditions or instruments. In particular, when the number of candidate moment conditions diverges, some conditions may be invalid or redundant, which motivates penalized GMM formulations for simultaneous moment selection and estimation \citep{41cheng2015select,42BelloniChernozhukovChetverikovHansenKato2018}. A related complication is that some candidate instruments may be invalid; in this setting, penalized IV procedures such as sisVIVE provide a useful regularized alternative to standard two-stage least squares \citep{kang2016instrumental}. More recently, computational issues in econometric GMM have also been revisited in online settings, where data arrive sequentially, and the estimator is updated recursively rather than recomputed from the full sample each time. In this direction, \citet{leung2025online} develops an online GMM framework with explicitly updated estimation, online weighting-matrix construction, and online versions of tests of overidentifying restrictions for streaming time-series data.

\subsection{Spatial statistics}
% DOMAIN 7: Spatial statistics

Estimating equations also arise naturally in spatial statistics when likelihood-based inference is computationally difficult. In clustered spatial point process models, such as Cox processes, the observed point pattern is generated through a latent random intensity structure, and likelihood-based inference is often computationally demanding and may require Monte Carlo methods \citep{moller2003statistical}. A computationally attractive alternative is to estimate the regression parameter $\beta$ through the estimating equation induced by the Poisson score. For inhomogeneous Neyman--Scott processes, this yields an unbiased estimating equation for $\beta$ even though the underlying process is not Poisson \citep{waagepetersen2007estimating}. However, because this approach relies only on first-order intensity information, it does not account for dependence among events and may therefore be statistically inefficient.

To improve efficiency, \citet{guan2010weighted} proposed a weighted estimating equation for inhomogeneous spatial point processes, where the weight function incorporates information on both inhomogeneity and dependence.  Building on this line of work, \citet{57thurman2015regularized} developed regularized methods for simultaneous estimation and variable selection in clustered spatial point processes.  The optimal weights proposed in \citet{guan2010weighted} depend on the coefficient vector $\beta$.  In \citet{57thurman2015regularized}, akin to the two-step GMM method, an initial estimate of $\beta$ is used to construct estimated weights, which are then fixed in the subsequent optimization program corresponding to the weighted estimating equation.  However, if we allow the weights to vary simultaneously with $\beta$ as in the continuously updating GMM estimator, then the weighted estimating equation no longer corresponds to the gradient of a scalar objective.

%In the weighted regularized formulation, the criterion depends on estimated first- and second-order structure through the weight function, so computation becomes iterative: one updates $\beta$ under the current weights, re-estimates the weight function, and then repeats the optimization until convergence.
\subsection{Outlier-prone data}
% Domain 8: Outlier-prone and contaminated data

Outliers and contamination provide another important source of non-standard structure in estimating equations. To reduce sensitivity to large residuals, robust M-estimators replace the square loss by a loss $\rho$, with derivative $\rho'$, that grows more slowly in the tails.  In linear regression, this leads to the estimating equation
\begin{equation}\label{eq:robust}
\bUn(\beta) \;=\; \frac{1}{n}\sum_{i=1}^{n} X_i\,\rho'(Y_i - X_i^\T\beta) \;=\; \mathbf 0.
\end{equation}
When $\rho'$ is monotone, as in Huber-type procedures, $\rho$ is convex, and the inversion of \eqref{eq:robust} becomes the empirical, convex loss $n^{-1} \sum_{i=1}^n \rho(Y_i-X_i^\T\beta)$, so standard penalized optimization remains applicable.  By contrast, when $\rho'$ is redescending, as in Tukey's bisquare or Hampel's proposal, $\rho$ becomes nonconvex, \eqref{eq:robust} may admit multiple roots, and the addition of a penalty further complicates the optimization landscape. In this sense, outlier-robust estimation leads naturally to estimating equations corresponding to both convex and nonconvex regularized optimization, depending on the choice of the loss $\rho$.
A recent example is \citet{28li2026penalized}, who combine a robust loss with nonconcave penalization in a penalized robust estimating-equation framework for semiparametric variable selection.
\subsection{Cross-domain}
\label{sec:EL}
% Empirical likelihood
Finally, empirical likelihood (EL) differs from the preceding examples in that it is not tied to a single application domain; rather, it provides a nonparametric likelihood framework that can be combined with any of the estimating functions discussed above.  We include it here because the computational structure of penalized EL is qualitatively different from those of the penalized versions of GEE, AFT, or GMM that we will review in Section~\ref{sec:algorithms}.  EL assigns observation-level weights $p_i$, so it is natural to work with the per-observation estimating function $\bU(W;\,\beta)\in\RR^q$ rather than the sample average $\bUn$: one maximizes $\sum_{i=1}^n \log p_i$ subject to $\sum_{i=1}^n p_i\,\bU(W_i;\,\beta) = \bfz$, $\sum_{i=1}^n p_i = 1$, and $p_i \ge 0$ \citep{owen1988empirical, qin1994empirical}.  Through Lagrange duality, the constrained problem reduces to an optimization over a multiplier $\eta\in\RR^q$, so for each fixed $\beta$ the EL objective is a smooth concave function of $\eta$.  Adding a sparsity penalty on $\beta$ then produces a two-layer problem: the inner layer optimizes over $\eta$ for a given $\beta$, while the outer layer solves a penalized problem in $\beta$ whose objective is only implicitly defined through the inner solution.  This nesting precludes a single-level penalized optimization and instead demands iterative procedures that alternate between the two layers \citep{52chang2018new, 53wang2019penalized, 56chen2023penalized, 55fu2024robust, 54tang2020penalized}.
\begin{table}[!t]
\centering
\caption{Estimating equation methods across settings and application domains, and their structural features discussed in this review.  The abbreviation EE stands for estimating equation.}
\label{tab:ee_domains}
\small
\renewcommand{\arraystretch}{1.2}
\begin{tabularx}{\textwidth}{@{} 
>{\raggedright\arraybackslash}p{2.7cm}
>{\raggedright\arraybackslash}p{2.7cm}
>{\raggedright\arraybackslash}p{3.0cm}
Y
@{}}
\toprule
\textbf{Setting / domain} & \textbf{EE framework} & \textbf{Foundational ref.} & \textbf{Structural feature} \\
\midrule
Longitudinal / clustered data & GEE, GEE2, ALR, QLS, QIF & Liang \& Zeger (1986); Qu et al.\ (2000) & Jacobian of $\bUn$ generally asymmetric; overidentified when using QIF \\[4pt]
Survival / censored data & Rank-based EE, Buckley--James & Tsiatis (1990); Buckley \& James (1979) & Piecewise-constant estimating function; Gehan weight admits convex loss but other weights do not \\[4pt]
Semiparametric models & Efficient scores & Bickel et al.\ (1993) & Efficient score obtained through projection, and does not correspond to a gradient \\[4pt]
Econometrics & IV, GMM & Hansen (1982) & Overidentified system \\[4pt]
Spatial statistics & Poisson surrogate EE, weighted EE & Waagepetersen (2007); Guan \& Shen (2010) & Full likelihood intractable; estimating equation depends on estimated weights \\[4pt]
Outlier-prone data & M-estimation EE & Li et al.\ (2026); Valdora \& Agostinelli (2025) & Redescending $\psi$ with nonconvex $\rho$ \\[4pt]
Cross-domain & EL with EE constraints & Owen (1988); Qin \& Lawless (1994) & Nested optimization over multiplier and parameter \\
\bottomrule
\end{tabularx}
\end{table}
% \begin{table}[!t]
% \centering
% \caption{Estimating equation methods across settings and application domains, their structural features discussed in this review.}\label{tab:ee_domains}
% \small
% \renewcommand{\arraystretch}{1.2}
% \begin{tabular}{@{} >{\raggedright}p{2.4cm} >{\raggedright}p{2.5cm} >{\raggedright}p{2.76cm} >{\raggedright\arraybackslash}p{5.2cm} @{}}
% \toprule
% \textbf{Setting / domain} & \textbf{EE framework} & \textbf{Foundational ref.} & \textbf{Structural feature} \\
% \midrule
% Longitudinal / clustered data & GEE, GEE2, ALR, QLS, QIF & Liang \& Zeger (1986); Qu et al.\ (2000) & Jacobian of $\bUn$ generally asymmetric; overidentified when using QIF \\[4pt]
% Survival / censored data & Rank-based EE, Buckley--James & Tsiatis (1990); Buckley \& James (1979) & Piecewise-constant estimating function; Gehan weight admits convex loss \\[4pt]
% Semiparametric models & Efficient scores & Bickel et al.\ (1993) & Efficient score depends on nuisance parameter through projection \\[4pt]
% Econometrics & IV, GMM & Hansen (1982) & overidentified system \\[4pt]
% Spatial statistics & Poisson surrogate EE, weighted EE & Waagepetersen (2007); Guan \& Shen (2010) & Full likelihood intractable; weights depend on estimated pair correlation \\[4pt]
% Outlier-prone data & M-estimation EE & Li et al.\ (2026); Valdora \& Agostinelli (2025) & Non-monotone $\psi$ with nonconvex $\rho$ \\[4pt]
% Cross-domain & EL with EE constraints & Owen (1988); Qin \& Lawless (1994) & Nested optimization over multiplier and parameter \\
% \bottomrule
% \end{tabular}
% \end{table}

\section{Computational algorithms}
\label{sec:algorithms}

What the preceding examples share is that, estimating equations often do not fit the standard optimization template.  Once a sparsity penalty is added to estimate $\beta^*$ and perform variable selection in high dimensions \citep{FanLiZhangZou2020}, where the dimension of $\beta^*$ could far exceed the sample size, the same problem can only aggravate in the resulting regularized estimating equations (REEs).  As we saw in Section~\ref{sec:EE_examples}, the specific culprits differ from case to case, including an asymmetric Jacobian in GEE, overidentification in QIF and GMM, piecewise-constant estimating functions in AFT rank regression, nonconvexity of the loss in robust M-estimation with redescending $\psi$, and a nested optimization structure in penalized EL.  Despite this diversity, the practical consequence is the same: widely used penalized optimization algorithms such as coordinate descent, iterative thresholding, and local linear approximation typically rely on the estimating function being the gradient of a scalar objective, and they do not directly apply when this structure is absent.

To date, penalized extensions of estimating equations have been developed largely within individual methodological communities, each proposing its own iterative scheme and deriving convergence guarantees under problem-specific assumptions.  We survey a number of popular existing approaches in Section~\ref{sec:previous}.  One caveat is that, given the vast number of areas where estimating equations can arise (as we have seen in Section~\ref{sec:EE_examples}), not all schemes are applicable to all problems (for instance, Section~\ref{sec:EL} most conveniently leads to a nested penalized optimization, not a regularized estimating equation).  The rest of this review takes a different approach.  By recognizing that a regularized estimating equation in the just-identified case (that is, when $q=p$) can be formulated as a fixed-point problem (FPP), one gains access to a family of algorithms whose convergence theory applies under conditions general enough to cover the structural features catalogued above.  We review this formulation, proposed in \cite{yang2021flexible}, in Section~\ref{sec:FPP}. 

\subsection{Previous methods for REEs}
\label{sec:previous}

Existing REE methods differ not only in how the estimating-equation problem is formulated, but also in the computational machinery used to solve the resulting problem. We therefore review the most commonly seen numerical strategies.

Let $\mathbf U_n(\beta)=(U_{n,1}(\beta),\ldots,U_{n,q}(\beta))^\top:
\mathbb R^p\to\mathbb R^q$ be an empirical estimating function based on a sample 
of size $n$, where $\beta=(\beta_1,\ldots,\beta_p)^\top\in\mathbb R^p$.  Thus \(q\) denotes the dimension of the estimating function output, whereas \(p\) denotes the dimension of the parameter of interest.  Existing REE formulations may be organized into three broad classes, namely 
minimization-type, Dantzig-type, and regularization-type, which we cover sequentially 
from Section~\ref{sec:min_REE} to Section~\ref{sec:reg_REE}.  Recall that 
we regard $\mathbf U_n(\cdot)$ as the sample analogue of the population estimating 
function $\mathbb E[\mathbf U_n(\cdot)]$, and the target parameter $\beta=\beta^*$ 
should be a well-isolated root to 
\(\mathbb E[\mathbf U_n(\beta)]=\mathbf 0\).   The minimization-type 
and Dantzig-type formulations (Sections~\ref{sec:min_REE} and \ref{sec:Dantzig_REE}) can accommodate possibly overidentified settings with \(q\ge p\). 
In contrast, the regularization-type REE formulation (Section~\ref{sec:reg_REE}) and the fixed-point reformulation 
are most directly defined in the just-identified case \(q=p\), when 
\(\mathbf U_n(\beta)\) has the same dimension as \(\beta\). To regularize estimation 
in high dimensions, let \(\Omega_\lambda(\beta)\) denote a penalty function, indexed 
by a tuning parameter \(\lambda>0\), used to induce sparsity or stabilize estimation.  The regularization-type 
formulation is arguably the most prominent, though it is also the one most closely 
tied to the just-identified case \(q=p\).

%########################################

\subsubsection{Minimization-type REEs}
\label{sec:min_REE}
In the minimization-type REE formulation, the estimating equation problem is first recast as an explicit penalized optimization problem of the form
\[
\hat\beta
\in
\arg\min_{\beta\in\mathbb R^p}
\Bigl\{
L_n(\beta)+\Omega_\lambda(\beta)
\Bigr\},
\]
where $L_n(\beta)$ is a scalar loss constructed from the estimating function.  A simple example with a coordinate-wise separable penalty $\Omega_\lambda$ is
\begin{align}
\label{eq:minimization_ee_1}
L_n(\beta)=\frac12\|\mathbf U_n(\beta)\|_2^2,
\qquad
\Omega_\lambda(\beta)=\sum_{j=1}^p p_\lambda(|\beta_j|),
\end{align}
where \(p_\lambda(\cdot)\) is a univariate penalty function, such as the \(\ell_1\) penalty for Lasso, or a nonconvex penalty. When \(L_n\) is more generally chosen as a weighted quadratic form of the estimating function, this construction yields a penalized GMM-type problem.  The defining feature of this class is that the estimator $\hat\beta$ is obtained as the minimizer of a penalized objective rather than as the root of a penalized estimating equation. Accordingly, computation is primarily optimization-oriented.

Within this class, several solver families appear repeatedly. When the transformed criterion is smooth or approximately smooth, the dominant tools are Newton--Raphson minimization and quasi-Newton methods, especially in penalized quadratic inference function (QIF), quasi-likelihood, or related formulations. This pattern appears in \citet{32cho2013model,35green2023semiparametric}, where the numerical core is a second-order optimization routine, possibly combined with additional approximation layers. When the objective involves nonconvex penalties such as smoothly clipped absolute deviation (SCAD), the local quadratic approximation (LQA) or local linear approximation (LLA) is often used to replace the original criterion by a sequence of simpler surrogate subproblems; see, for example, \citet{21he2020high,32cho2013model,35green2023semiparametric}. When the penalized objective has a coordinate-wise or block-wise structure, as is often the case in high-dimensional problems, coordinate descent or nested coordinate descent provides a natural computational strategy. This is especially common in the empirical-likelihood and generalized empirical-likelihood literature, including \citet{52chang2018new,53wang2019penalized,54tang2020penalized,55fu2024robust,56chen2023penalized}, where the algorithm typically alternates between updating the model parameter and updating the associated Lagrange multiplier. A smaller set of papers invokes more specialized solvers tailored to the objective at hand, such as threshold gradient descent regularization in \citet{22tian2021gee}, projected scaled sub-gradient active-set updates in \citet{41cheng2015select}, LP and exact Lasso-path computation in \citet{47cai2009regularized}, quasi-Newton schemes such as limited-memory BFGS in \citet{48chung2013tutorial}, and penalty-specific combinations of least angle regression (LARS), one-step LLA, and coordinate descent in \citet{57thurman2015regularized}. Thus, once the estimating-equation problem has been converted into a scalar objective, the computational task becomes one of structured penalized optimization, and the numerical method is chosen according to the geometry of that objective.

We would like to point out that the minimization carried out in a minimization-type REE formulation should not be placed on the same footing as a typical minimization task with a ``natural'' origin (for instance, one directly based on the negative log-likelihood).   Note that the loss function $L_n$ in a minimization-type REE formulation, which corresponds to a zeroth-order condition, is built from $\mathbf U_n(\beta)$ which can be thought of as a first-order condition.  Thus this $L_n$ is typically more complex than those arising from a natural minimization problem.  For instance, in a simple linear regression problem, $\mathbf U_n(\beta) = \frac{1}{n} \sum_{i=1}^n X_i (Y_i - X_i^\top \beta) \equiv  \frac{1}{n} \mathbf{X}^\top(\textbf{Y}-\mathbf{X}\beta)$ where $\mathbf{X}\in\RR^{n\times p}$ is the covariate matrix formed by stacking the individual predictors $X_i^\top$ vertically and $\textbf{Y}$ is formed by stacking the responses vertically.  In this case, it is easy to see that the Hessian of the loss $L_n$ in \eqref{eq:minimization_ee_1} is $( \frac{1}{n} \mathbf{X}^\top \mathbf{X})^2$, whose condition number is the \textit{square} of that in a direct least squares problem. 
This illustrates that minimization-type reformulations may be less well-conditioned than the corresponding native optimization problems, which can affect numerical performance.

%########################################

\subsubsection{Dantzig-type REEs}
\label{sec:Dantzig_REE}
A second line of work does not begin with an explicit penalized loss. Instead, it seeks a sparse parameter vector that approximately satisfies the estimating equation by directly constraining the estimating-function norms (\citet{johnson2011path, yu2013dimension,li2014dantzig, LeeBrzyskiBogdan2016, shi2018high, 42BelloniChernozhukovChetverikovHansenKato2018, choiruddin2023adaptive}). A representative formulation is
\[
\hat\beta
\in
\arg\min_{\beta\in\mathbb R^p}\Omega_\lambda(\beta)
\qquad\text{subject to}\qquad
\Omega_\lambda^*( \mathbf U_n(\beta) ) \le \tau,
\]
where $\Omega_\lambda^*$ denotes the dual norm of $\Omega_\lambda$ and $\tau>0$ controls the tolerated deviation from zero of the empirical estimating function $\mathbf U_n$.  In the most common special case, $\Omega_\lambda(\beta)=\lambda\|\beta\|_1$, yielding the familiar Dantzig selector-type formulation.  Then, one guiding principle for choosing $\tau$ is that it should be large enough such that the target $\beta^*$ satisfies the constraint, which in this case becomes $\lambda \|\mathbf U_n(\beta^*)\|_\infty \le \tau$ where $\|\cdot\|_\infty$ denotes the elementwise maximum norm, with high probability.  The essential feature of this formulation is that sparsity is induced through the objective, while approximate satisfaction of the estimating equation is imposed through a sup-norm feasibility constraint. This viewpoint appears in a range of settings, including censored regression \citep{johnson2011path, li2014dantzig}, semiparametric dimension-reduction problems \citep{yu2013dimension}, estimating-equation-based treatment-regime and spatial-intensity problems \citep{shi2018high, choiruddin2023adaptive}, and more general constrained Dantzig-type formulations with structured penalties \citep{LeeBrzyskiBogdan2016}.

%$\mathbf A_n(\beta)$ is constructed from the estimating function $\mathbf U_n(\beta)$ or moment condition, possibly after projection, standardization, or other transformation, and

The dominant computational tools in this branch are LP and related constrained optimization methods, especially when the estimating function is linear or can be accurately approximated by a linear form. In such cases, the Dantzig problem can often be written as a standard linear program, sometimes together with a dual formulation, making large-scale computation relatively tractable. This computational structure is particularly clear in \citet{li2014dantzig}, where a censored linear regression problem is reformulated through an imputed Buckley--James-type estimating equation and then solved by an adaptive Dantzig selector, and in \citet{LeeBrzyskiBogdan2016}, where a generalized Dantzig selector with an ordered $\ell_1$ penalty is handled through a saddle-point algorithm for the resulting constrained problem. When the estimating function is nonlinear, however, standard LP is no longer directly applicable, and an additional linearization step is usually required before solving a sequence of tractable constrained subproblems.  To conduct inference on a low-dimensional parameter constructed from $\beta$ (for instance, a fixed set of coordinates of $\beta$), the Dantzig-type REE formulation can again be invoked in a further projection or correction step; see, for example, \citet{42BelloniChernozhukovChetverikovHansenKato2018} and \citet{61NeykovNingLiuLiu2018} as well as the subsequent, more general debiased machine learning setup \cite{chernozhukov2018double}. 
Group Lasso penalties also require separate treatment. Although they can be incorporated into constrained or Dantzig-type formulations, the groupwise $\ell_2$ norms do not generally lead to a standard linear program. Such formulations are more naturally handled by second-order cone programming, block-coordinate methods, or proximal algorithms, and therefore should not be identified with the LP-based Dantzig-type REE formulation considered here.
\subsubsection{Regularization-type REEs}
\label{sec:reg_REE}

In a regularization-type REE formulation, the estimating equation structure is preserved, and the penalty (required to be coordinate-wise separable) enters directly at the equation level. In this setting, the regularized estimating equation typically seeks the root $\hat\beta$ of
\begin{equation}
\label{eq:old_ee}
    \mathbf U_n(\beta)+\mathbf q_\lambda(|\beta|)\odot\beta=\mathbf 0,
\end{equation}
where
$\mathbf q_\lambda(|\beta|)
=
\bigl(q_\lambda(|\beta_1|),\dots,q_\lambda(|\beta_p|)\bigr)^\top$, $\odot$ denotes componentwise multiplication, and $\lambda>0$ is a tuning parameter controlling the amount of regularization.  As the form of \eqref{eq:old_ee} clearly suggests, the regularization-type REE formulation is only applicable in the just-identified, $q=p$, case.  Starting from a coordinate-wise separable penalty of the form
\[
\sum_{j=1}^p \rho_\lambda(|\beta_j|) ,
\]
with $\rho_\lambda'$ denoting the derivative of $\rho_\lambda$, one typically sets
\[
q_\lambda(|\beta_j|)=\frac{\rho_\lambda'(|\beta_j|)}{|\beta_j|},
\]
so that
\[
q_\lambda(|\beta_j|)\beta_j
=
\rho_\lambda'(|\beta_j|)\sgn(\beta_j),
\qquad \beta_j\neq 0.
\]
Hence, unlike a minimization-type REE formulation, the estimator remains defined as the root of a penalized estimating equation rather than as the minimizer of an explicit scalar objective.

The existing regularization-type literature can be organized more clearly by its computational backbone. A first major group consists of papers that solve the penalized estimating equation through iteratively reweighted least-squares or local-surrogate Newton-type updates. \citet{fu2003penalized} is an early representative, where penalization is incorporated directly into generalized estimating equations through a penalized iteratively reweighted least squares (IRLS) procedure. \citet{15johnson2008variable} develops the same direct root-solving idea using SCAD penalization together with LQA and Newton--Raphson / Fisher-scoring-type updates. In the same computational family, \citet{16wang2012penalized} considers SCAD-penalized generalized estimating equations (GEEs) for high-dimensional longitudinal data under a diverging-$p$ framework, \citet{17inan2017pgee} provides a software implementation of penalized GEE using cyclic coordinate descent together with LQA, and \citet{18geronimi2017variable} extends penalized GEE to multiply-imputed longitudinal data by combining multiple imputation, the local quadratic approximation of the group Lasso penalty, and a modified Newton--Raphson algorithm. Thus, these papers belong to the same broad solver family in which the penalty is handled locally and the resulting penalized equation is solved by IRLS- or Newton-type iteration.

A second major group consists of papers that enrich this basic root-solving template by majorization--minimization (MM), iterative weighted least squares (IWLS), or related surrogate-based refinements. The underlying idea is still to solve the penalized estimating equation directly, but the nonconvex or nonsmooth penalty is first replaced by a tractable local surrogate before each update. \citet{16wang2012penalized} already has this flavor through its SCAD treatment; \citet{19zhang2018model} formulates an adaptive Lasso-penalized robust generalized estimating equation system and solves it by IWLS combined with MM, \citet{20lin2020group} uses a Newton--Raphson plus MM algorithm for group-penalized GEE with structured correlation, and \citet{59yang2020doubly} remains in the same family by combining folded-concave penalization with MM and coordinate-wise Newton--Raphson updates. In this sense, these papers may be viewed as a common MM-based regularization-type subfamily, differing mainly in the penalty structure and in whether the final update is global Newton, weighted least squares, or coordinate-wise Newton.

A third group consists of papers that preserve the regularization-at-the-equation-level formulation but incorporate more structured numerical devices tailored to the underlying model. \citet{23ma2025penalized} studies penalized weighted generalized estimating equations under informative cluster size and computes the estimator by a quasi-Newton / Newton--Raphson algorithm with active-set updating. \citet{27cruz2024penalized} considers penalized generalized estimating equations for repeated-measures crossover designs with complex carry-over effects and solves the modified penalized estimating equations by an iterative block-wise scheme that alternates updates of functional coefficients, carry-over coefficients, fixed effects, and working correlation parameters. \citet{28li2026penalized} develops a penalized robust estimating equation in a semiparametric model and combines local linear smoothing, kernel-based bias correction, reparametrization, and Fisher-scoring / Newton--Raphson iteration. These papers therefore form a structured-iteration subfamily of regularization-type REEs, in which the direct penalized root-solving framework is retained but supplemented by active-set strategies, block-wise iteration, smoothing, or bias-correction layers.

Taken together, the common feature across these papers is that the penalty is inserted directly into the estimating equation, and the estimator is obtained by iterative root-solving rather than by reformulating the problem as an unconstrained penalized minimization problem. The main computational differences lie not in the overall formulation, but in the numerical mechanism used to stabilize and accelerate the iterations: IRLS, LQA, MM, IWLS, cyclic coordinate descent, active-set updating, block-wise iteration, and smoothing-based correction.

A particularly influential template in this branch is the LQA strategy underlying papers \citet{15johnson2008variable,16wang2012penalized,17inan2017pgee,18geronimi2017variable}. Specifically, given a current anchor $\tilde\beta_j$, one approximates
\[
\rho_\lambda(|\beta_j|)
\approx
\rho_\lambda(|\tilde\beta_j|)
+
\frac12
\frac{\rho_\lambda'(|\tilde\beta_j|)}{|\tilde\beta_j|}
\bigl(\beta_j^2-\tilde\beta_j^2\bigr),
\qquad
\beta_j\approx \tilde\beta_j,
\]
which yields
\[
\frac{\partial}{\partial\beta_j}\rho_\lambda(|\beta_j|)
=
\rho_\lambda'(|\beta_j|)\sgn(\beta_j)
\approx
\frac{\rho_\lambda'(|\tilde\beta_j|)}{|\tilde\beta_j|}\beta_j.
\]
Under this approximation, the penalized estimating equation is locally replaced by
\[
Q_{\tilde\beta}(\beta)
\equiv
\mathbf U_n(\beta)+\boldsymbol{\Lambda}_\lambda(\tilde\beta)\odot\beta
=
\mathbf 0,
\]
where
\[
\boldsymbol{\Lambda}_\lambda(\tilde\beta)
=
\diag\!\left\{
\frac{\rho_\lambda'(|\tilde\beta_1|)}{|\tilde\beta_1|},
\dots,
\frac{\rho_\lambda'(|\tilde\beta_p|)}{|\tilde\beta_p|}
\right\}.
\]
The next iterate is then computed by
\[
\beta^{(k+1)}
=
\beta^{(k)}
-
\left[
\frac{\partial \mathbf U_n(\beta^{(k)})}{\partial\beta^\top}
+
\boldsymbol{\Lambda}_\lambda(\beta^{(k)})
\right]^{-1}
Q_{\beta^{(k)}}(\beta^{(k)}).
\]
This LQA--Newton template is common, but it also has clear limitations. 
Because the update requires inverting a $p\times p$ Jacobian-plus-penalty matrix at each iteration, the computational cost can become substantial in high dimensions. 
Moreover, the local quadratic surrogate is tailored mainly to coordinate-wise penalties and is less natural for more general structured penalties, such as group Lasso or SCAD. 
The iteration also does not directly generate sparsity; in practice, additional thresholding or ad hoc stabilization is often needed, especially near zero coefficients. 
These limitations help motivate later reformulations of REE through fixed-point mappings and related first-order methods.

\subsection{Fixed-point equivalence}
\label{sec:FPP}

While much of the earlier literature treats regularized estimating equations via direct root-finding, often via LQA- or Newton-type updates, these methods can become restrictive in high dimensions and are not always well-suited to nonsmooth or structured penalties.  A more recent line of work therefore emphasizes that the same problem can be reformulated as a fixed-point problem \citep{yang2021flexible}. This perspective is useful not only algorithmically, but also conceptually, because it provides a unified way to handle general nonsmooth convex and nonconvex penalties, including structured penalties.

To make this connection precise, it is helpful to begin with a subdifferential formulation rather than the superficially similar classical coordinate-wise penalized Eq.~\eqref{eq:old_ee} discussed in in Section~\ref{sec:reg_REE}. For simplicity, we only describe the method under the convex setting.  Given a convex and possibly nonsmooth penalty $\Omega_\lambda:\mathbb R^p\to\mathbb R$, we seek the solution $\hat\beta$ to the regularized estimating equation formulated as
\begin{align}
0 \in \mathbf U_n(\beta) + \partial \Omega_\lambda(\beta),
\label{eq:fpp_ee}
\end{align}
where $\mathbf U_n(\beta)$ is the estimating function and $\partial\Omega_\lambda(\beta)$ denotes the subdifferential of $\Omega_\lambda$ at $\beta$. This form is more general than the usual coordinate-wise representation in Eq.~\eqref{eq:old_ee} earlier, since it is not restricted to coordinate-wise separable $\Omega_\lambda$ and remains valid for structured convex penalties such as the group Lasso and sparse group Lasso.  Compared to Eq.~\eqref{eq:old_ee}, the formulation in \eqref{eq:fpp_ee} also relaxes the condition on $\mathbf U_n$ at those coordinates $j\in\{1,\dots,p\}$ where $\hat\beta_j$ is zero.  For example, under the Lasso penalty $\Omega_\lambda(\beta)=\lambda\|\beta\|_1$, the subgradient condition allows any inactive coordinate $j$ to satisfy $[\mathbf U_n(\hat\beta)]_j\in[-\lambda,\lambda]$, rather than forcing a pointwise equality. In this sense, the subdifferential form provides a more complete characterization of the regularized estimating equation than the original coordinate-wise expression \eqref{eq:old_ee}.

The fixed-point connection arises by observing that, for any $\tau>0$ that can be regarded as a step size, the inclusion \eqref{eq:fpp_ee} at the solution $\beta=\hat\beta$, namely
\[
0 \in \mathbf U(\hat\beta) + \partial \Omega_\lambda(\hat\beta),
\]
can be equivalently written as the first-order optimality condition of the strongly convex problem
\[
\hat\beta
=
\arg\min_{\beta\in\mathbb R^p}
\left\{
\frac{1}{2}
\left\|
\beta-\bigl(\hat\beta-\tau \mathbf U(\hat\beta)\bigr)
\right\|_2^2
+
\tau\Omega_\lambda(\beta)
\right\} ;
\]
see Section~3 in \cite{yang2021flexible} for the complete derivation.  Introducing the proximal operator
\[
\operatorname{prox}_{\Omega}(v)
=
\arg\min_{z}
\left\{
\frac{1}{2}\|z-v\|_2^2+\Omega(z)
\right\}
\]
for a generic penalty $\Omega$.  One then obtains the equivalent fixed-point representation
\[
\hat\beta=f(\hat\beta),
\qquad\text{where}\qquad
f(\beta)=\operatorname{prox}_{\tau\Omega_\lambda}\bigl(\beta-\tau \mathbf U_n(\beta)\bigr).
\]
Thus, solving the convex regularized estimating equation is equivalent to finding a fixed-point of the map $f$ formed by an estimating-equation update followed by a proximal regularization step. In the unregularized case $\lambda=0$, the proximal map reduces to the identity, so the same construction recovers the usual fixed-point formulation of the nonregularized estimating equation. From the viewpoint of this review, this equivalence is the key bridge linking convex REE formulations with fixed-point problems.

This reformulation is also computationally informative.  Evaluating the proximal operator for a convex penalty amounts to solving a simple strongly convex subproblem, and for many commonly used penalties, the proximal map is available in closed form. For the Lasso, for instance, the proximal map reduces to the soft-thresholding rule
\[
[\operatorname{prox}_{\tau\lambda\|\cdot\|_1}(v)]_j
=
\operatorname{sgn}(v_j)(|v_j|-\tau\lambda)_+.
\]
Here, $(a)_+ = \max\{a,0\}$ denotes the positive part, and
$[\cdot]_j$ denotes the $j$-th component of the argument vector. Thus the
proximal map is evaluated componentwise: coordinates with
$|v_j|\le \tau\lambda$ are set to zero, whereas coordinates with
$|v_j|>\tau\lambda$ are shrunk toward zero by $\tau\lambda$.
Accordingly, sparsity is generated directly by the proximal step itself, rather than being imposed afterward through an auxiliary thresholding rule. More generally, structured penalties such as the group Lasso and sparse group Lasso also admit efficient proximal evaluations, making the fixed-point form natural beyond the coordinate-wise setting.

Once the REE has been written in this way, several iterative schemes become available. The most direct one is the Picard iteration \citep{picard1890memoire,banach1922operations}
\[
\beta^{(k+1)}
=
\operatorname{prox}_{\tau\Omega_\lambda}
\bigl(\beta^{(k)}-\tau \mathbf U_n(\beta^{(k)})\bigr),
\]
which repeatedly applies the fixed-point map. A more stable alternative is the relaxed Krasnosel'skii--Mann iteration \citep{mann1953mean,krasnosel1955two}
\[
\beta^{(k+1)}
=
(1-\rho)\beta^{(k)}
+
\rho\,
\operatorname{prox}_{\tau\Omega_\lambda}
\bigl(\beta^{(k)}-\tau \mathbf U_n(\beta^{(k)})\bigr),
\qquad \rho\in(0,1).
\]
When the fixed-point map \(f\) defined earlier is contractive 
\citep[Section~3]{ryu2016primer}, the Picard iteration converges geometrically to a unique 
fixed point. Under the weaker nonexpansiveness assumption on $f$, the Krasnosel'skii--Mann 
iteration provides a standard, relaxed scheme for approaching a fixed point and for obtaining 
bounds on the fixed-point residual.

From a broader computational perspective, the fixed-point equivalence has several advantages. As already mentioned, it accommodates general convex penalties through proximal mappings and therefore is not limited to coordinate-wise separable penalties. However, the scope of this viewpoint extends beyond the convex case, as \citet{yang2021flexible} notes that the equivalence between 
regularized estimating equations and fixed-point problems also holds for more general penalty functions, including nonconvex penalties. Second, each update is first-order in nature and avoids the large matrix inversions that often arise in Newton-type or LQA methods. Third, the resulting framework connects REE computation to the broader literature on fixed-point algorithms, proximal methods, and monotone operator theory \citep{ryu2016primer}, thereby allowing one to draw on scalable iterative procedures and existing convergence theory. For this reason, fixed-point reformulations are best viewed not merely as an alternative notation, but as a flexible computational route for solving regularized estimating equations.

\section{Conclusion}
\label{sec:conclusion}

This review examined the computation of regularized estimating equations from a formulation-based perspective. We considered four broad formulations: minimization-type, Dantzig-type, regularization-type, and fixed-point-type, and reviewed the main numerical strategies associated with each of them. The central theme is that different representations of a regularized estimating equation lead to different computational approaches for solving it.

%A main conclusion of the review is that
%The traditional computation of REEs is often closely tied to the structural form of the underlying problem.

Across application domains, the main difficulties associated with (regularized) estimating equations may arise from asymmetric Jacobians, overidentified moment systems, nonsmooth estimating functions, nonconvexity, or nested optimization.  Traditionally, these structural features often determine the representation of the resultant regularized estimating equation, which in turn dictates which computational tools are applicable and what theoretical analysis could follow. From this perspective, the connection between regularized estimating equations and fixed-point problems places a broad range of existing structural forms within a common computational framework.

Several directions for future work remain open, including scalable first-order algorithms for very high-dimensional problems, and sharper convergence guarantees under weaker structural assumptions.   It is also important to develop computational frameworks that more directly accommodate nuisance estimation, complex dependence, and model-specific constraints. On the practical side, clearer benchmarking studies and more accessible software implementations would improve the comparability and clarify the most applicable scenarios of existing REE methods.

\section*{Author Contribution}
\textbf{Weihua Shi}: investigation (equal), visualization (equal), writing -- original draft (equal), \textbf{Yixuan Li}: investigation (equal), visualization (equal), writing -- original draft (equal), \textbf{Yi Lian}: software (lead), \textbf{Archer Y. Yang}: conceptualization (equal), supervision (equal), validation (equal), writing -- review and editing (equal), \textbf{Yue Zhao}: conceptualization (equal), supervision (equal), validation (equal), writing -- review and editing (equal)

\section*{Funding Information}
This work was partially supported by CANSSI Collaborative Research Teams Grant, NSERC Discovery Grant (RGPIN-2024-06780) and FRQNT Team Research Project Grant (FRQNT 327788).

\section*{Conflicts of Interest}
The authors declare no conflicts of interest.

\section*{Acknowledgements}

The authors acknowledge the use of OpenAI ChatGPT (GPT-5.5 Thinking, accessed April 2026) in manuscript preparation, specifically to assist with literature review and language polishing.  AI use did not impact key arguments or conclusions.  The authors independently verified all suggestions, revised the text as needed, and take full responsibility for the final content.

\noindent This work was done in part while Yue Zhao was visiting IVADO.

\section*{Data Availability}
Data sharing is not applicable to this article as no new data were created or analyzed in this study.

\bibliography{eeLasso}

\end{document}